**Homogeneity in the instrument-treatment association is not sufficient for the Wald estimand to equal the average causal effect for a binary instrument and a continuous exposure**

Fernando Pires Hartwig[1,2]*, Linbo Wang[3], George Davey Smith[2,4], Neil Martin Davies[2,4,5]

[1]Postgraduate Program in Epidemiology, Federal University of Pelotas, Pelotas, Brazil.

[2]MRC Integrative Epidemiology Unit, University of Bristol, Bristol, United Kingdom.

[3]Department of Statistical Sciences, University of Toronto, Toronto, ON, Canada.

[4]Population Health Sciences, Bristol Medical School, University of Bristol, Bristol, United Kingdom.

[5]K.G. Jebsen Center for Genetic Epidemiology, Department of Public Health and Nursing, NTNU, Norwegian University of Science and Technology, Norway.

*Corresponding author. Postgraduate Program in Epidemiology, Federal University of Pelotas, Pelotas (Brazil). Adress: Rua Marechal Deodoro 1160 (3rd floor). ZIP code: 96020-220. Phone: 55 53 981126807. E-mail: fernandophartwig@gmail.com

**ABSTRACT**

**Background:** Interpreting instrumental variable results often requires further assumptions in addition to the core assumptions of relevance, independence, and the exclusion restriction.

**Methods:** We assess whether instrument-exposure additive homogeneity renders the Wald estimand equal to the average derivative effect (ADE) in the case of a binary instrument and a continuous exposure.

**Results:** Instrument-exposure additive homogeneity is insufficient for ADE identification when the instrument is binary, the exposure is continuous and the effect of the exposure on the outcome is non-linear on the additive scale. For a binary exposure, the exposure-outcome effect is necessarily additive linear, so the homogeneity condition is sufficient.

**Conclusions:** For binary instruments, instrument-exposure additive homogeneity identifies the ADE if the exposure is also binary. Otherwise, additional assumptions (such as additive linearity of the exposure-outcome effect) are required.

**BACKGROUND**

Valid inference from instrumental variable (IV) analysis requires that the candidate instrument is valid – that is, it satisfies the three core IV assumptions of relevance (the instrument is – preferably strongly – statistically associated with the exposure or treatment variable), independence (there are no unmeasured common causes between the instrument and the outcome) and exclusion restriction (instrument and outcome are independent conditional on the exposure and all exposure-outcome confounders).[1,2] Moreover, point estimation of well-defined causal estimands requires additional assumptions, sometimes referred to as IV4 assumptions.[3] Examples of IV4 assumptions include homogeneity of the association between the instrument and the exposure and homogeneity of the causal effect of the exposure on the outcome. These two assumptions (on the additive scale) are often postulated to render the conventional Wald estimand to equal the average causal effect (ACE). Both of these assumptions are special cases of a more general assumption named "NO Simultaneous Heterogeneity" (NOSH).[4] Another prominent IV4 assumption is that of monotonicity in the instrument-exposure association, in which case the Wald estimand equals a conditional causal estimand – the ACE among compliers (also known as the local ACE).[5]

Additive homogeneity in the instrument-exposure association has been extensively demonstrated to be a sufficient IV4 assumption for identifying the ACE for a binary exposure.[6,7] Indeed, since the conventional Wald estimand can be interpreted as a weighted average of individual-level causal exposure-outcome effects, where the weights stem from the individual-level instrument-exposure association, it is rather intuitive that instrument-exposure additive homogeneity is also sufficient when the exposure is continuous.

However, when the exposure is continuous, the exposure-outcome effect is not necessarily linear on the additive scale, which introduces additional complexities. If the effect is non-linear, the ACE will depend on the pair of exposure values being compared, which leads to several (indeed, infinite in the case of continuous exposures) ACE parameters.[3] So, there would be no single ACE, and the question of whether the Wald estimand equals the ACE would just be ill-defined.

Conversely, the Wald estimand is defined for a continuous exposure, and understanding whether it equals a well-defined causal estimand of practical interest is useful not only for proper interpretation of the results, but also for the practical relevance of the method. One possible way of summarizing the causal effect of a continuous exposure is the average derivative effect (ADE)[8,9] (defined in detail below). This parameter is often referred to as the continuous analogue of the ACE, because it (approximately) quantifies the effect of an intervention that increases the exposure by one unit in everyone in a given population; and because it is equivalent to the ACE for binary exposures (because the effect of binary exposures is necessarily additive linear).

Although the ADE may be a natural extension of the ACE for continuous exposures, it is not necessarily the case that IV4 assumptions sufficient for ACE identification are also sufficient for ADE identification. Here we demonstrate that additive homogeneity in the association between the instrument and the exposure is not sufficient for the ADE to be identified even if the three core IV assumptions hold when the instrument is binary and the exposure is continuous.

**METHODS**

Let $Z$ denote the instrument, $X$ the exposure, $Y$ the outcome and $U$ any common causes of $X$ and $Y$. We assume that $Z$ is a valid causal instrument, that is, $Z$ has a causal effect on $X$ and $Z$ satisfies the three core

IV assumptions of relevance ($Z \not\!\perp\!\!\!\perp X$), independence ($Y^{(Z=z)} \perp\!\!\!\perp Z$, where $Y^{(Z=z)}$ denote the potential outcome under a hypothetical intervention that sets $Z$ to some fixed value $z$) and exclusion restriction ($Z \perp\!\!\!\perp Y|X,U$).

We assume $X$ and $Y$ are generated by the following non-parametric structural equation model:

$$X = f_X(Z, U, \varepsilon_X)$$

$$Y = f_Y(X, U, \varepsilon_Y).$$

In this model, $f_X$ and $f_Y$ denote generic functions respectively governing $X$ and $Y$, which we assume to be differentiable with respect to $Z$ and $X$, respectively. Moreover, $\varepsilon_X$ and $\varepsilon_Y$ (where $\varepsilon_X \perp\!\!\!\perp \varepsilon_Y$) respectively denote stochastic direct causes of $X$ and $Y$.

We consider the case where $Z$ is binary and $X$ is continuous. The ADE can be defined as $\text{ADE} = \text{E}\left[\frac{\partial Y}{\partial X}\right] = \text{E}\left[\frac{\partial}{\partial X} f_Y(X, U, \varepsilon_Y)\Big|_{X=X_i, U=U_i, \varepsilon_Y=\varepsilon_{Y_i}}\right]$. Therefore, the derivative causal effect for individual $i$ can be interpreted as the partial derivative function evaluated at $X_i$, $U_i$ and $\varepsilon_{Y_i}$, and the ADE is simply the average of such individual level effects in the population under study (because the expectation is taken over the population-specific joint distribution of $X$, $U$ and $\varepsilon_Y$).

For a binary $Z$, the conventional Wald estimand is $\beta_{IV} = \frac{\text{E}[\beta_{Z,Y}]}{\text{E}[\beta_{Z,X}]}$, where $\beta_{Z,X}$ and $\beta_{Z,Y}$ respectively denote the individual-level difference in $X$ and $Y$ according to $Z$ (i.e, $X^{(Z=1)} - X^{(Z=0)}$ and $Y^{(Z=1)} - Y^{(Z=0)}$, respectively).[5]

Initially we use theoretical arguments to support our claims. After that, we illustrate the conclusions with an example.

**RESULTS**

Since $Z$ is binary, $f_X$ is necessarily linear on the additive scale with respect to $Z$. This implies the following:

$$f_X = c_X Z + Z g_{1_X}(U, \varepsilon_X) + g_{2_X}(U, \varepsilon_X) + \alpha_X \Rightarrow \frac{\partial X}{\partial Z} = c_X + g_{1_X}(U, \varepsilon_X).$$

$$\beta_{Z,X} = X^{(Z=1)} - X^{(Z=0)}$$

$$= \left[c_X + g_{1_X}(U, \varepsilon_X) + g_{2_X}(U, \varepsilon_X) + \alpha_X\right] - \left[g_{2_X}(U, \varepsilon_X) + \alpha_X\right]$$

$$= c_X + g_{1_X}(U, \varepsilon_X) = \frac{\partial X}{\partial Z}.$$

We initially assume that $f_Y$ is linear on the additive scale with respect to $X$ (a condition that is guaranteed to hold if $X$ is binary). Therefore:

$$f_Y = c_Y X + X g_{1_Y}(U, \varepsilon_Y) + g_{2_Y}(U, \varepsilon_Y) + \alpha_Y \Rightarrow \frac{\partial Y}{\partial X} = c_Y + g_{1_Y}(U, \varepsilon_Y).$$

$$\beta_{Z,Y} = Y^{(Z=1)} - Y^{(Z=0)}$$

$$= \left(c_Y X^{(Z=1)} + X^{(Z=1)} g_{1_Y}(U, \varepsilon_Y) + g_{2_Y}(U, \varepsilon_Y) + \alpha_Y\right)$$
$$- \left(c_Y X^{(Z=0)} + X^{(Z=0)} g_{1_Y}(U, \varepsilon_Y) + g_{2_Y}(U, \varepsilon_Y) + \alpha_Y\right)$$
$$= c_Y \left[X^{(Z=1)} - X^{(Z=0)}\right] + g_{1_Y}(U, \varepsilon_Y)\left[X^{(Z=1)} - X^{(Z=0)}\right]$$
$$= \left(X^{(Z=1)} - X^{(Z=0)}\right)\left(c_Y + g_{1_Y}(U, \varepsilon_Y)\right) = \frac{\partial X}{\partial Z}\frac{\partial Y}{\partial X}.$$

From the above,

$$\beta_{IV} = \frac{\mathrm{E}[\beta_{Z,Y}]}{\mathrm{E}[\beta_{Z,X}]} = \frac{\mathrm{E}\left[\frac{\partial X}{\partial Z}\frac{\partial Y}{\partial X}\right]}{\mathrm{E}\left[\frac{\partial X}{\partial Z}\right]}.$$

If $\beta_{Z,X}$ and $\beta_{X,Y}$ are independent (which is essentially what the NOSH assumption postulates), then:

$$\frac{\mathrm{E}\left[\frac{\partial X}{\partial Z}\frac{\partial Y}{\partial X}\right]}{\mathrm{E}\left[\frac{\partial X}{\partial Z}\right]} = \frac{\mathrm{E}\left[\frac{\partial X}{\partial Z}\right]\mathrm{E}\left[\frac{\partial Y}{\partial X}\right]}{\mathrm{E}\left[\frac{\partial X}{\partial Z}\right]} = \mathrm{E}\left[\frac{\partial Y}{\partial X}\right] = \mathrm{ADE}.$$

That is, when $Z$ is binary and $X$ is continuous, the conventional Wald estimand equals the ADE if the effect of $X$ on $Y$ is additive linear and $\beta_{Z,X}$ and $\beta_{X,Y}$ are independent.

We now consider the case where the effect of $Z$ on $X$ is homogeneous on the additive scale. This implies the following:

$$f_X = c_X Z + \alpha_X \implies \frac{\partial X}{\partial Z} = c_X.$$

$$\beta_{Z,X} = X^{(Z=1)} - X^{(Z=0)} = [c_X + \alpha_X] - [\alpha_X] = c_X = \frac{\partial X}{\partial Z}.$$

Therefore, if $f_Y$ is linear on the additive scale with respect to $X$, $\beta_{IV} = \mathrm{ADE}$ because $\beta_{Z,X}$ and $\beta_{X,Y}$ are necessarily independent given $Z$-$X$ additive homogeneity.

However, $\beta_{IV} = \mathrm{ADE}$ does not necessarily hold if $f_Y$ is non-linear on the additive scale with respect to $X$. To understand why this is the case, assume $\beta_{Z,Y} = \frac{\partial Y}{\partial X}\left(X^{(Z=1)} - X^{(Z=0)}\right)$. In this case:

$$\beta_{IV} = \frac{\mathrm{E}[\beta_{Z,Y}]}{\mathrm{E}[\beta_{Z,X}]} = \frac{\mathrm{E}\left[\frac{\partial Y}{\partial X}\left(X^{(Z=1)} - X^{(Z=0)}\right)\right]}{\mathrm{E}[X^{(Z=1)} - X^{(Z=0)}]} = \frac{\left(X^{(Z=1)} - X^{(Z=0)}\right)\mathrm{E}\left[\frac{\partial Y}{\partial X}\right]}{X^{(Z=1)} - X^{(Z=0)}} = \mathrm{ADE}$$

Therefore, if $\beta_{Z,Y} = \frac{\partial Y}{\partial X}\left(X^{(Z=1)} - X^{(Z=0)}\right)$, then $\beta_{IV} = \mathrm{ADE}$.

For a binary $Z$, $\beta_{Z,Y} = Y^{(X^{(Z=1)})} - Y^{(X^{(Z=0)})} = \frac{Y^{(X^{(Z=1)})} - Y^{(X^{(Z=0)})}}{X^{(Z=1)} - X^{(Z=0)}}\left[X^{(Z=1)} - X^{(Z=0)}\right]$, that is, the average slope between the points $(X^{(Z=0)}, Y^{(X^{(Z=0)})})$ and $(X^{(Z=1)}, Y^{(X^{(Z=1)})})$ multiplied by $X^{(Z=1)} - X^{(Z=0)}$. Therefore, the equality $\beta_{Z,Y} = \frac{\partial Y}{\partial X}\left(X^{(Z=1)} - X^{(Z=0)}\right)$ is satisfied if $\frac{\partial Y}{\partial X} = \frac{Y^{(X^{(Z=1)})} - Y^{(X^{(Z=0)})}}{X^{(Z=1)} - X^{(Z=0)}}$. Since $X =$ $X^{(Z=0)}$ or $X = X^{(Z=1)}$, $\frac{\partial Y}{\partial X}$ is the partial derivative at one of the endpoints of the interval $\left[X^{(Z=0)}, X^{(Z=1)}\right]$ (if $X^{(Z=0)} < X^{(Z=1)}$) or $\left[X^{(Z=1)}, X^{(Z=0)}\right]$ (if $X^{(Z=0)} > X^{(Z=1)}$). Since the average slope between two points is only guaranteed to equal the derivative at one of the endpoints of the interval if the curve is a line, $\beta_{IV} = $ ADE is only guaranteed to hold under $Z$-$X$ additive homogeneity if $f_Y$ is linear on the additive scale with respect to $X$.

To illustrate the theoretical conclusions above, consider the following example:

- $Z \sim \text{Bernoulli}(0.3)$
- $U \sim \text{Bernoulli}(0.5)$
- $X = 2Z + U + \varepsilon_X$, where $\varepsilon_X \sim N(0,1)$
- $Y = 2X^2 + U + \varepsilon_Y$, where $\varepsilon_Y \sim N(0,1)$

Under this data-generating model:

$$E[X] = E[2Z + U + \varepsilon_X] = 1.1$$

$$\text{ADE} = E\left[\frac{\partial}{\partial X}[2X^2 + U + \varepsilon_Y]\right] = 4E[X] = 4.4$$

$$\beta_{Z,X} = X^{(Z=1)} - X^{(Z=0)} = 2 = \frac{\partial X}{\partial Z}.$$

$$\frac{\partial Y}{\partial Z} = \frac{\partial}{\partial Z}[2(2Z + U + \varepsilon_X)^2 + U + \varepsilon_Y]$$

$$= 8(2Z + U + \varepsilon_X) \Rightarrow E[\beta_{Z,Y}] = 8.8.$$

$$\beta_{Z,Y} = Y^{(Z=1)} - Y^{(Z=0)}$$

$$= (2(2 + U + \varepsilon_X)^2 + U + \varepsilon_Y) - (2(U + \varepsilon_X)^2 + U + \varepsilon_Y)$$

$$= 8(1 + U + \varepsilon_X) \Rightarrow E[\beta_{Z,Y}] = 12 \neq E\left[\frac{\partial X}{\partial Z}\right]E\left[\frac{\partial Y}{\partial X}\right].$$

From the above, $\beta_{IV} = \frac{12}{2} = 6 \neq \text{ADE}$.

**DISCUSSION**

Both our theoretical arguments and the numerical example show that, contrary to a notion that is present in the literature, $Z$-$X$ additive homogeneity is not sufficient in general for the Wald estimand to equal the ADE when $Z$ is binary and $X$ is continuous even if $Z$ is a valid causal instrument. As shown above, additionally assuming the effect of $X$ on $Y$ is linear on the additive scale allows identification of the ADE.

Although this is a rather technical point, understanding the conditions required for interpreting IV estimates as being consistent for some well-defined estimand (in this case, the ADE) is important for proper interpretation of the results, which otherwise are just numbers with no clear meaning and therefore of little practical use. Indeed, the importance that empirical studies explicitly acknowledge and justify not only the assumptions required for the candidate instrument to be valid, but also for point identification of well-defined estimands, is clearly recognized, for example, in reporting guidelines for instrumental variable analyses.[10,11]

The present results may also have practical implications for IV analysis. For example, empirical approaches for assessing $Z$-$X$ additive homogeneity based on variance comparisons have recently been proposed.[12,13] The rationale for this would be that, if there is no strong evidence against homogeneity, then it would be plausible to interpret the results as estimates of the ACE. Such rationale itself appears to implicitly assume additive linearity in the effect of $X$ on $Y$, otherwise the ACE is not uniquely defined. If the ADE is used as the definition of the ACE for a continuous $X$, our conclusions demonstrate that, even if one could prove that $Z$-$X$ additive homogeneity holds, this would still not be sufficient to identify the ADE when $Z$ is binary. This implies that, even if such strong homogeneity assumptions holds, interpreting the Wald estimate without yet further assumptions (such as additive linearity in the effect of $X$ on $Y$) may be challenging.

**ACKNOWLEDGEMENTS**

The Medical Research Council (MRC) and the University of Bristol support the MRC Integrative Epidemiology Unit [MC_UU_00011/1/2]. NMD is supported by a Norwegian Research Council Grant number 295989. LW is partially supported by a McLaughlin Accelerator Grant in Genomic Medicine.

This work is dedicated to the memory of Mr. Dari Hartwig (FPH's father).